\documentclass[
  aps,
  prl,
reprint,
  amsmath, amssymb,
  ]{revtex4-2}
\usepackage[T1]{fontenc}
\usepackage{mathptmx}
\usepackage[
]{xcolor}
\definecolor{linkblue}{RGB}{31,119,180}
\usepackage[
  unicode,
  colorlinks,
  citecolor=blue,
  linkcolor=linkblue,
  urlcolor=linkblue,
  bookmarks=true,
  bookmarksopen=true,
  bookmarksopenlevel=3,
  bookmarksnumbered=true]{hyperref}
\usepackage{siunitx}
\usepackage{graphicx}
\usepackage{amsmath}
\usepackage{amssymb}
\usepackage{textcomp}
\usepackage{listings}
\usepackage{courier}
\usepackage{longtable}
\usepackage{booktabs}

\graphicspath{{img/}}

\newcommand{\gsw}{gravitational-wave}

\newcommand{\sqz}{\SI{7.2(2)}{dB}}
\newcommand{\asqz}{\SI{15.6(2)}{dB}}

\newcommand{\qe}{\SI{92(3)}{\%}}

\newcommand{\pump}{\SI{1064}{nm}}
\newcommand{\convert}{\SI{2128}{nm}}

\begin{document}

\preprint{AIP/123-QED}

\title{Squeezed light at 2128 nm for future gravitational-wave observatories}

\newcommand{\ilp}{Institut für Laserphysik und Zentrum für Optische Quantentechnologien der Universität Hamburg, Luruper Chaussee 149, 22761 Hamburg, Germany}

\author{Christian Darsow-Fromm}
    \email{christian.darsow@physik.uni-hamburg.de}
    \affiliation{\ilp}
\author{Julian Gurs}
    \email{julian.gurs@physik.uni-hamburg.de}
    \affiliation{\ilp}
\author{Roman Schnabel}
    \affiliation{\ilp}
\author{Sebastian Steinlechner}
    \email{s.steinlechner@maastrichtuniversity.nl}
    \affiliation{Department of Gravitational Waves and Fundamental Physics, Maastricht University, P.O. Box 616, 6200 MD Maastricht, The Netherlands}
    \affiliation{Nikhef, Science Park 105, 1098 XG Amsterdam, The Netherlands}

\begin{abstract}
    All gravitational-wave observatories (GWOs) have been using the laser wavelength of \pump{}.
Ultra-stable laser devices are at the sites of GEO\,600, Kagra, LIGO and Virgo.
Since 2019, not only GEO\,600 but also LIGO and Virgo have been using separate devices for squeezing the uncertainty of the light, so-called squeeze lasers.
The sensitivities of future GWOs will strongly gain from reducing the thermal noise of the suspended mirrors, which involves shifting the wavelength into the \SI{2}{\micro m} region.
Our work aims for reusing the existing high-performance lasers at 1064\,nm.
Here, we report the realisation of a squeeze laser at \convert{} that uses ultra-stable pump light at \pump{}.
We achieve the direct observation of \SI{7.2}{dB} of squeezing, as the first step, at MHz sideband frequencies.
The squeeze factor achieved is mainly limited by the photodiode's quantum efficiency, which we estimated to \qe.
Reaching larger squeeze factors seems feasible, also in the required audio and sub-audio sideband, provided photo diodes with sufficiently low dark noise will be available.
Our result promotes \convert{} as the new, cost-efficient wavelength of GWOs. 

 \end{abstract}

\maketitle

\section{Introduction}
\label{sec:introduction}

The third observing run of LIGO and VIRGO \gsw\ observatories (GWOs) produced a plethora of varied and unique astrophysics events, limited by fundamental noise sources \cite{abbottGW190814GravitationalWaves2020}. 
GWOs with a tenfold increased reach for sources producing signal frequencies around \SI{100}{Hz} and with hundred times larger range around \SI{10}{Hz} seem feasible \cite{ETSteeringCommitteeEditorialTeam2020,Reitze2019}.
Such high sensitivities will expand the detection range toward the entire universe for some sources, will result in a quasi permanent observation of mutually overlapping signals, and will promise new insights into cosmology and even the origin of the universe.

Current GWOs are limited by residual seismic noise, control noise and photon radiation pressure noise at sub-audio frequencies, by thermally excited internal movement of the mirror coatings (coating thermal noise) in the lower audio-band, and by photon counting noise in the higher audio-band \cite{ligoscientificcollaborationInstrumentScienceWhite2019}.
Changing the laser wavelength from \pump{} to around \SI{2}{\micro m} will allow the usage of crystalline silicon as the bulk material of the test mass mirrors that are cryogenically cooled to about 18\,K, potentially in combination with high-quality silicon-based coatings \cite{steinlechnerSiliconBasedOpticalMirror2018a}.
Increasing the signal requires ultra-stable laser radiation, that is not absorbed or scattered by the test mass mirrors. 
Reducing the quantum noise of the radiation requires squeezing the optical quantum uncertainty \cite{cavesQuantummechanicalNoiseInterferometer1981,schnabelQuantumMetrologyGravitational2010,schnabelSqueezedStatesLight2017} over the entire spectrum of expected signals, as first achieved in \cite{vahlbruchQuantumEngineeringSqueezed2007}.
Current GWOs use well-proven ultra-stable laser devices with powers of up to \SI{160}W and squeeze lasers with a nonclassical noise suppression between  \SI7{dB} and \SI{12}{dB} \cite{vahlbruchGEO600Squeezed2010,mehmetHighefficiencySqueezedLight2018}.
Optical resonators increase the light powers to up to \SI{750}{kW} in the \SI4{km} arms in the case of Advanced LIGO \cite{TheLIGOScientificCollaboration2015}.
Optical loss reduces the squeeze factor to \SI6{dB} in the case of GEO\,600 \cite{theligoscientificcollaborationGravitationalWaveObservatory2011,groteFirstLongTermApplication2013,loughFirstDemonstrationDB2021} and around \SI3{dB} in LIGO and Virgo \cite{tseQuantumEnhancedAdvancedLIGO2019,acerneseIncreasingAstrophysicalReach2019}.

A first squeeze laser for the \SI{2}{\micro m} region was previously reported in \cite{yapSqueezedVacuumPhase2019a,mansellObservationSqueezedLight2018}.
Squeeze factors of up to \SI{4}{dB} were measured.
The value was limited by the quantum efficiency of the photo detectors, as well as noise of the \SI{1984}{nm} thulium fiber laser and subsequently its second harmonic pump field at \SI{992}{nm}.

Here, we report the realisation of a squeeze laser at \convert\ that uses ultra-stable \pump\ pump light from a Nd:YAG nonplanar ring oscillator (NPRO), which is also used as the master laser in current GWOs.
We directly observed a squeeze factor of \sqz{} at sideband-frequencies around \SI{2}{MHz}.
The squeezed field uncertainty was observed by a balanced homodyne detector that used a bright stable local oscillator beam at \convert\ that was produced by degenerate optical parametric oscillation (DOPO), which we reported previously~\cite{darsow-frommHighlyEfficientGeneration2020}.

 \section{Experimental Setup}\label{experimental-setup}

\begin{figure}
    \includegraphics[width=\linewidth]{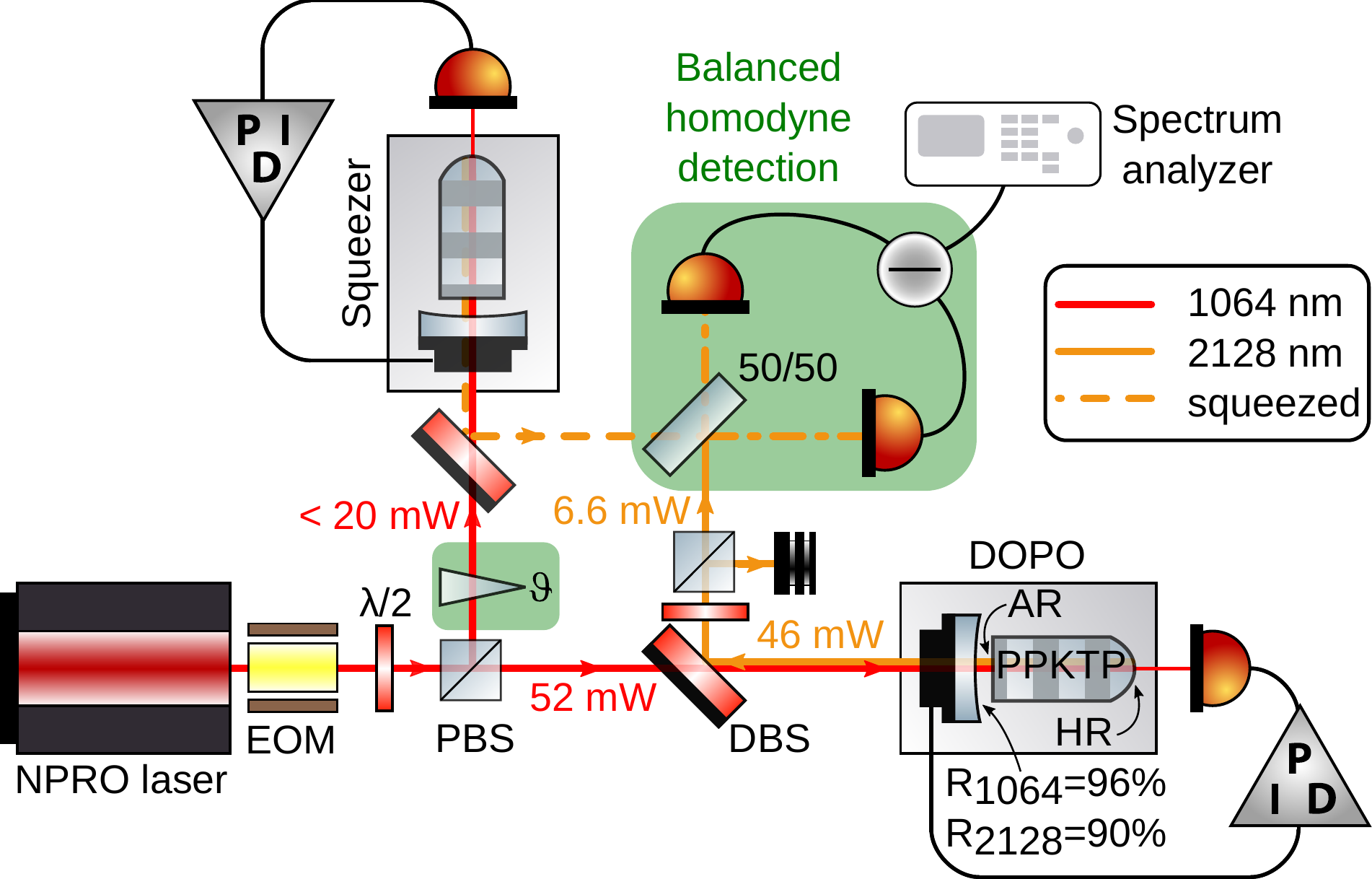}
    \caption{
        Schematics of the experiment.
        The NPRO laser provided up to \SI{2}{W} output power at the wavelength $\lambda = \SI{1064}{nm}$ for the two identical nonlinear cavities: squeezer and DOPO. The squeeze light was detected with a balanced-homodyne detector. EOM, electro-optical modulator; PBS, polarizing beam-splitter; DBS, dichroic beam-splitter.
    }
    \label{fig:lasersystem}
\end{figure}

Our experimental setup (Fig.~\ref{fig:lasersystem}) is based on two identical nonlinear resonators that are optimized for wavelength-doubling via degenerate optical-parametric amplification.
The two resonators were pumped with continuous-wave 1064\,nm light from an NPRO laser.
The resonators had a half-monolithic (hemilithic) design and were composed of periodically-poled potassium titanyl phosphate (PPKTP) crystals, with highly-reflective coating on the curved end face and an anti-reflective coating on the flat front face, and separate coupling mirrors with reflectivities of \SI{96}\% at \pump\ and \SI{90}\% at \convert.
An electro-optical modulator provided phase-modulation sidebands at \SI{28}{MHz} for a modified Pound-Drever-Hall control scheme in transmission of the resonators together with a digital controller \cite{darsow-frommNQontrolOpensourcePlatform2020} to stabilize the length of the DOPO cavity on resonance by feeding back to the piezo-mounted coupling mirrors.
One resonator was pumped above the lasing threshold for DOPO (around \SI{20}{mW}) and produced about \SI{46}{mW} at \convert{} from about \SI{52}{mW} at \pump{}; detailed information on this setup can be found in a previous publication \cite{darsow-frommHighlyEfficientGeneration2020}.
The other resonator was pumped below the threshold power and therefore produced a well defined light beam with a TEM\textsubscript{00} mode in a squeezed vacuum state at 2128\,nm.

The generated squeezing was analyzed with a balanced-homodyne detector (BHD).
For this, it was overlapped with a bright beam from the DOPO on a 50\% beam splitter and both outputs were sent to photodiodes (extended InGaAs, Thorlabs FD05D), whose photocurrents were then subtracted from each other.
The readout angle of the balanced-homodyne detector could be adjusted with a phase shifter, i.e.\ a piezo-mounted mirror, that was located in front of the squeezed-light cavity to suppress induced pointing loss.
Our self-made electronics operated the photodiodes at a reverse bias voltage of \SI{1}{V} and achieved a detection bandwidth of \SI{30}{MHz}.
The quantum efficiency of the extended-InGaAs photodiodes slightly increases with higher reverse voltage, but the dark-current and noise rises rapidly.
We have found a reverse voltage of \SI1V to be a useful balance between quantum efficiency and noise.
With the photodiodes' windows removed, we measured a quantum efficiency of \qe\ with a thermal power-meter (accuracy \SI3\%) and precise multimeters.
 \section{Results}\label{results}

\begin{figure*}
    \includegraphics[width=\linewidth]{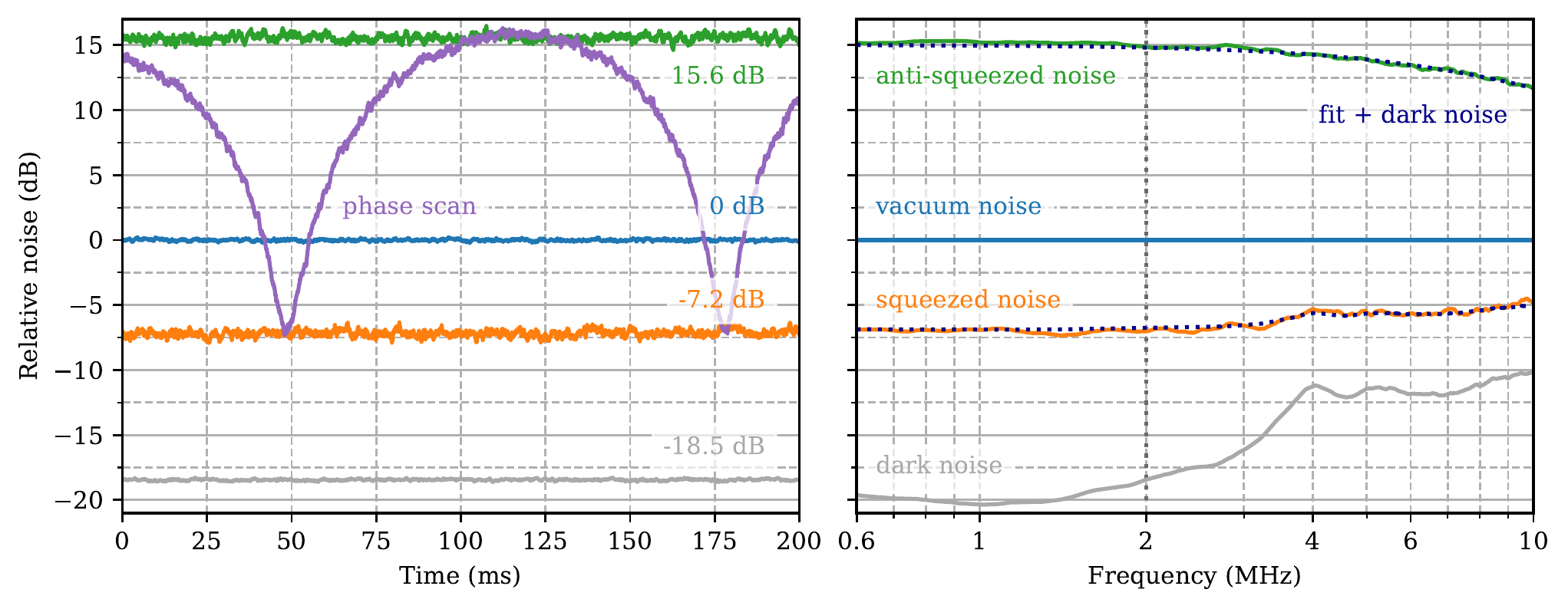}
    \caption{
        (Left)  
        Zero-span noise measurement at a sideband frequency of \SI{2}{MHz}.
        We achieved a squeezed noise reduction of \sqz\ below the vacuum noise, accompanied with an anti-squeezed noise in the orthogonal quadrature of \asqz.
        The noise arches were obtained by scanning the BHD readout angle.
        All traces were recorded with a resolution bandwidth of \SI{300}{kHz} and a video bandwidth of \SI{300}{Hz}.
        Dark noise and vacuum noise were additionally averaged 10 times.
        \\
        (Right) 
        Spectrum of the generated squeeze light in the regime \SIrange{0.6}{10}{MHz}, fitted with the equations~(\ref{eq:phasenoise}) and (\ref{eq:sqz_spectrum}), where the dark noise was added to the fitting curves.
        All traces were averaged 10 times.
    }
    \label{fig:squeezing}
\end{figure*}

Fig.~\ref{fig:squeezing} presents noise variances from BHD measurements recorded with a spectrum analyzer, normalized to the vacuum noise. 
The left panel shows a zero-span measurement at 2\,MHz, while the right panel shows the spectrum in the range \SIrange{0.6}{10}{MHz}.
Electronic dark-noise was not subtracted from these traces.
We obtained a non-classical noise suppression of \sqz\ at a sideband frequency of \SI{2}{MHz} and a local-oscillator power of \SI{6.6}{mW}.
This squeezing level extended to lower frequencies, as shown in Fig.~\ref{fig:squeezing}\,(right), before the dark-noise clearance quickly decreased as the BHD's transfer function was optimized for the MHz regime. 
At our measurement frequency, electronic dark-noise was dominated by the dark-current of the photodiodes.
It was not subtracted from the measurements and reduced the achieved noise suppression by about \SI{0.3}{dB}.
We estimate the error on the squeezed/anti-squeezed noise levels to $\pm \SI{0.2}{dB}$, as the BHD readout angle was not yet servo controlled and therefore prevented longer measurements at the optimal quadratures.

\begin{figure}
    \includegraphics[width=\linewidth]{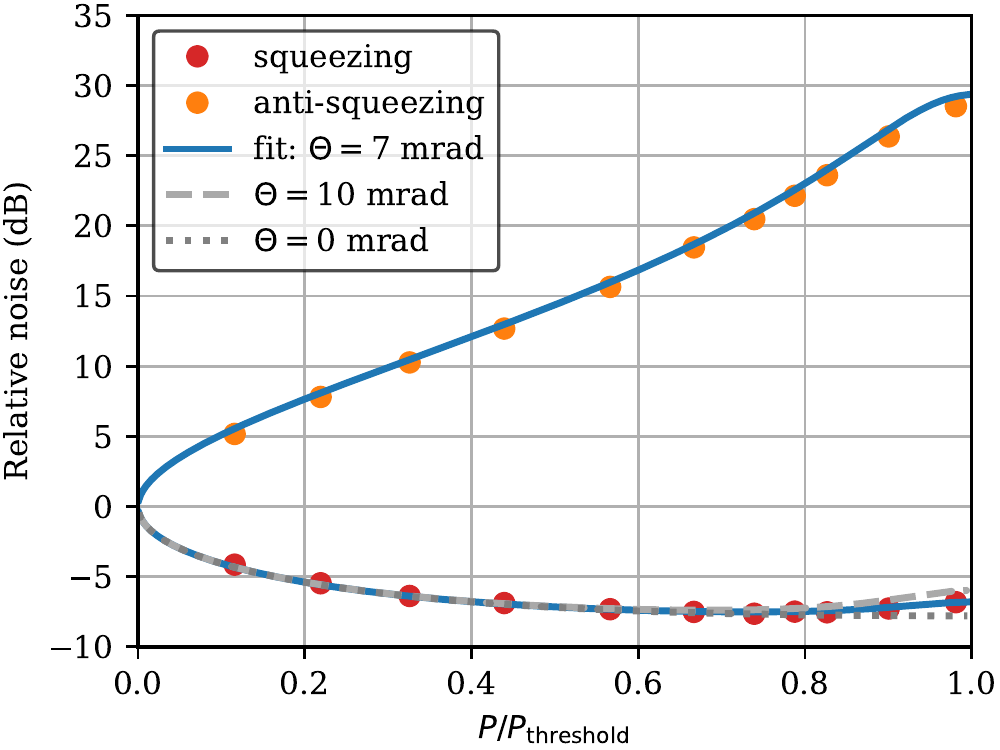}
    \caption{ 
        Dependence of squeezed and anti-squeezed noise levels on the pump power, as a fraction of threshold power.
        The data points were taken by using max hold for anti-squeezing and min hold for squeezing, comparing to the respective max/min hold vacuum noise reference.
        At high pump powers, the observed squeeze noise level degraded due to phase noise.
    }
    \label{fig:phase_noise}
\end{figure}

Random fluctuation of the phase between the squeezed-light beam and the local oscillator in the setup leads to a coupling between the squeezed and anti-squeezed light-field quadratures, which we denote here as $\hat X_1$ and $\hat X_2$, respectively.
For a small amount of gaussian-distributed phase noise with an rms value of $\Theta$, the measured quadrature variances $\Delta^2 \hat X_{1,2}^m$ are given by \cite{takenoObservationOf9dB2007}
\begin{equation}
    \Delta^2 \hat X_{1,2}^m = \Delta^2\hat X_{1,2} \, \cos^2\Theta
                                + \Delta^2\hat X_{2,1}\, \sin^2\Theta\,.
    \label{eq:phasenoise}
\end{equation}
As phase noise becomes particularly relevant for large variances of the anti-squeezed quadrature, an upper bound can be determined by a measurement of the squeezing and anti-squeezing levels for various pump powers $P$ up to the threshold power $P_\text{thr} = \SI{20}{mW}$.
The quadrature variances themselves can be described by \cite{takenoObservationOf9dB2007}
\begin{equation}
    \Delta^2 \hat{X}_{1,2} = 1 \mp \eta \frac{4\sqrt{ P / P_{\rm thr}}} {\bigl(1 \pm \sqrt{ P/P_{\rm thr} } \bigr)^2 + 4 (\Omega/\gamma)^2}\,,
    \label{eq:sqz_spectrum}
\end{equation}
where the upper sign corresponds to $\hat X_1$ and the lower sign to $\hat X_2$.
Here, the variance of the vacuum ground state has been normalized to 1; $\eta$ is the overall detection efficiency; $\gamma = 2\pi\times\SI{64}{MHz}$ is the linewidth of our squeezed-light cavity; and $\Omega = 2\pi\times\SI{2}{MHz}$ is the measurement sideband frequency.
Combining equations \eqref{eq:phasenoise} and \eqref{eq:sqz_spectrum}, we fitted our measurements Fig.~\ref{fig:phase_noise} and obtained an rms phase noise  of $\Theta = \SI{7(1)}{mrad}$.
This phase noise is likely dominated by high-frequency fluctuations introduced by the locking loops of the cavities, as well as residual high-frequency fluctuations of the main laser beam.
However, it is not limiting our squeezing results and we therefore did not yet implement steps to reduce it.

\paragraph{Optical loss analysis}
State of the art squeeze lasers are entirely limited by optical loss, which arises from absorption, scattering, imperfect mode matching and imperfect quantum efficiency of the photodiodes.
The total optical efficiency $\eta$ can be derived from a combination of the squeezed and anti-squeezed variances $\Delta^2 \hat X_1$ and $\Delta^2 \hat X_2$ (with dark noise subtracted),
\begin{equation}
    1-\eta = \frac{1-\Delta^2 \hat X_1 \Delta^2 \hat X_2}{2 - \Delta^2 \hat X_1 - \Delta^2 \hat X_2}\,.
\end{equation}
Inserting the measured values of \SI{-7.2}{dB} squeezing and \SI{15.6}{dB} anti-squeezing, we arrive at a an efficiency of $\eta = \SI{83.9(5)}\%$.

We have estimated the loss contributions in our setup from the measured quantum efficiency of the photodiodes and manufacturers' specifications of the optics. 
These are summarized in Tab.~\ref{tab:efficiency}.

\begin{table}[htb]
    \centering \vspace{-2mm}
    \caption{Overview of optical efficiencies}
    \label{tab:efficiency}
    \begin{tabular}{lc}
        \toprule
        Source  & Efficiency (\%) \\
        \midrule
        Resonator escape efficiency & \SI{98(1)}{} \\
        Propagation efficiency      & $>\SI{99}{}$ \\
        BHD visibility \SI{98(1)}{\%} & $\SI{96(2)}{}$ \\
        Photodiode quantum efficiency & \SI{92(3)}{} \\
        \midrule
        Total value as product of estimated efficiencies & \SI{85(4)}{} \\
        \midrule
        Total value from squeeze and anti-squeeze values & \SI{83.9(5)}{} \\
        \bottomrule
    \end{tabular}
\end{table}

The escape efficiency of the squeeze resonator is given by $T/(T+L)$, where $T$ is the coupling-mirror transmissivity, and $L$ is a sum of all round-trip losses, such as from an imperfect anti-reflective coating on the crystal, scattering and absorption loss, as well as residual transmission through the non-perfect reflecting back face of the crystal.

Propagation loss towards the homodyne detector is likely small, due to the use of high-quality optics and infrared-grade fused silica substrates, and has been estimated to \SI{<0.1}\% per surface.
The beam overlap (\emph{visibility}) between the local oscillator and squeezed beam at the balanced-homodyne detector contributes quadratically, and therefore has a high impact.
We measured a visibility of $V = \SI{98(1)}{\%}$.
Finally, we include the non-perfect quantum efficiency of our photodiodes, around \SI{92}\% according to our measurements, in the estimate.
This is the therefore the largest individual contribution.

Within its relatively large error bars, our estimated value for the overall efficiency is in agreement with the one obtained from the squeezing and anti-squeezing measurement.

 \section{Conclusion}

We have reported on a novel approach to combine squeezed light generation at \convert{} via parametric down-conversion with degenerate optical parametric oscillation pumped by an ultra-stable Mephisto laser at \pump{}.
We currently reach a squeeze level of \sqz{} in the MHz sideband frequencies, being mainly limited by the quantum efficiency of the available photodiodes.
The concept of wavelength-doubling, combined with squeezing, makes the wavelength \convert{} a promising, because cost-efficient candidate for all next-generation \gsw{} detectors like Cosmic Explorer \cite{Reitze2019}, Einstein Telescope \cite{ETpathfinderDesignReport2020}, NEMO \cite{Ackley2020} and Voyager \cite{adhikariCryogenicSiliconInterferometer2020}.
In these GWOs, a squeeze level of \SI{10}{dB} is usually aimed at.
A reduction of optical loss within the detector to around \SI{6.3}{\%} may be within reach, for realistic technological advances \cite[tab.~6.1]{Schreiber2017}.
The squeeze light source itself will then need to produce a measured squeeze level of \SI{15}{dB}, which has been demonstrated at a wavelength of \SI{1064}{nm} \cite{vahlbruchDetection15DB2016a}.
Further research into low-noise photo detectors with a quantum efficiency of \SI{99}\% is required to achieve this goal also at \SI{2}{\micro m}.

\begin{acknowledgments}
\vspace{3ex}
This research has been funded by the Deutsche Forschungsgemeinschaft (DFG, German Research Foundation) -- 388405737.

We acknowledge financial support by the Germany Federal Ministry of Education and Research, grant no. 05A20GU5.\\
This article has LIGO document number \href{https://dcc.ligo.org/P2100175}{P2100175}.
\end{acknowledgments}

\bibliography{references}

\end{document}